\documentclass[12pt]{article}

\usepackage{amssymb}
\usepackage{amsmath}
\usepackage{amsfonts}

\usepackage{graphicx}
\usepackage{xypic}

\begin{document}
\title{Ternary Poisson algebra for the non degenerate three dimensional Kepler Coulomb potential\footnote{Contribution to the 4th Workshop on Group Analysis of Differential Equations
and Integrable Systems, Protaras, Cyprus, Oct. 2008}}
\author{ Y. Tanoudis\thanks{tanoudis@math.auth.gr} and C. Daskaloyannis\thanks{daskalo@math.auth.gr} \\
Mathematics Department\\
Aristotle University of Thessaloniki\\
54124  Thessaloniki- Greece}
\date{Jan. 2009}

\maketitle

\begin{abstract}
In the three dimensional flat space any classical Ha\-mi\-lton\-ian, which  has five functionally independent integrals of motion, including the Hamiltonian, is characterized as superintegrable. Kalnins, Kress and Miller \cite{KalKrMi07} have proved that, in the case of non degenerate potentials, i.e potentials depending linearly on four parameters, with quadratic symmetries, posses a sixth quadratic integral, which is linearly independent of the other integrals. The existence of this sixth integral imply that the  integrals of motion form  a ternary parafermionic-like quadratic Poisson algebra with five generators. The Kepler Coulomb potential that was introduced by Verrier and Evans \cite{VerEvans08} is a special case of superintegrable system, having two independent integrals of motion of fourth order among the remaining quadratic ones.
The corresponding Poisson algebra of integrals is  a quadratic one, having the same special form, characteristic to the non degenerate case of systems with quadratic integrals.
\end{abstract}

\section{Introduction}\label{int1}

In classical mechanics, a superintegrable or completely integrable
is a Hamiltonian system with a maximum number of integrals. Two well
known examples are the harmonic oscillator and the Coulomb
potential. In the $N$-dimensi\-onal space the superintegrable system
has $2 N-1$ integrals, one among them is the Hamiltonian.

Several cases of three dimensional superintegrable systems with quadratic integrals of motion are described and analyzed by Kalnins, Kress and Miller \cite{KalKrMi07,KalKrMi07JPA}. Specifically, Kalnins, Kress and Miller studied a special case of superintegrable systems in which the potentials
depend of four free parameters, these systems are  referred  as  \emph{non degenerate potentials}.
In the case that one three dimensional potential have fewer arbitrary constants than four the potential is called \emph{degenerate}. The generate and non degenerate potentials have been studied by N.W. Evans \cite{Evans90}. One among the degenerate systems is the so called \emph{Gene\-ralized Kepler - Coulomb} system \cite{KaWiMiPo99,Evans90}
\begin{equation}\label{eq:Generalized-Coulomb}
H=\frac{1}{2} (p_{x}^2+p_{y}^2+p_{z}^2)-\dfrac{k }{ \sqrt{x^2+y^2+z^2}}+\dfrac{k_1}{x^2}+\dfrac{k_2}{y^2}
\end{equation}
This potential has four integrals of motion quadratic in momenta plus the Hamiltonian. These quadratic integrals of motion do not appear to close under repeated commutation and they do not satisfy a quadratic algebra\cite{KaWiMiPo99} as it happens for the supeintegrable two dimensional systems\cite{Das01}. Therefore one of the open problems is to find the Poisson algebra of the integrals of motion for the superintegrable system (\ref{eq:Generalized-Coulomb}).

One of the  results of the Kalnins, Kress, Miller paper \cite{KalKrMi07} is  the so called "5 to 6" Theorem, which states that any three dimensional non degenerate superintegrable system with quadratic integrals of motion has always a sixth quadratic integral $F$ that is linearly independent but not functionally dependent regarding the set of five integrals $A_1,A_2,B_1,B_2,H $. The last statement leads to the result that any three dimensional superintegrable non degenarate system form a, quadratic,  ternary  Poisson algebra of special character\cite{TaDas09}, whose the definition is given in Section \ref{para}.

Verier and Evans\cite{VerEvans08} introduced a new superintegrable Hamiltonian
\begin{equation}\label{eq:Non_degenerate}
H=\frac{1}{2} (p_{x}^2+p_{y}^2+p_{z}^2)-\frac{k }{ \sqrt{x^2+y^2+z^2}}+\frac{k_1}{x^2}+\frac{k_2}{y^2}+\frac{k_3}{z^2}
\end{equation}
which is the non degenerate version of the potential of the generalized Kepler - Coulomb system (\ref{eq:Generalized-Coulomb}).
The above potential is indeed superintegrable with quadratic and quartic in momenta integrals of motion. The quartic integrals are generalizations of the Laplace–
Runge–Lenz vectors of the ordinary Kepler - Coulomb potential\cite{VerEvans08}.

In this paper we prove that the "5 to 6" theorem of Kalnins, Kress, Miller\cite{KalKrMi07} can be applied and the associate Poisson algebra is a ternary quadratic algebra of the constants of motion, which are different of the Hamiltonian. This algebra is similar to the ternary parafermionic-like algebra for the three dimensional non degenerate potentials\cite{TaDas09}.
Therefore the algebra of the generalized Kepler - Coulomb system is also a ternary parafermionic-like algebra.

\section{Ternary Parafermionic-like Poisson Algebra}\label{para}

The definition of the Lie algebra  $\mathfrak{g}$ with generators $x_1,x_2,\ldots,x_n$ leads to the following relations
\[
[x_i,x_j]= \sum\limits_{m}c_{ij}^m x_m
\]
where $c_{ij}^m$ the structure constans.
The generators satisfy the obvious ternary (trilinear) relations
\begin{equation}\label{eq:DefLieTriple}
T\left(x_i,x_j,x_k\right) \mathop{\equiv}\limits^{\rm def } [x_i,[x_j,x_k]]= \sum\limits_{n} d_{i;jk}^n x_n, \quad
\mbox{ where } \; d_{i;jk}^n = \sum\limits_{m} c_{i m}^n c_{jk}^m
\end{equation}
Generally a ternary algebra is an associative algebra $\mathcal{A}$ satisfying whose the generators satisfy relations like the following one
\[
T\left(x_i,x_j,x_k\right) = \sum\limits_{n} d_{i;jk}^n x_n
\]
where $T\, :\, \mathcal{A}\otimes\mathcal{A}\otimes\mathcal{A} \longrightarrow \mathcal{A}$ is a trilinear map. If this trilinear map is defined as in eq. (\ref{eq:DefLieTriple}) the corresponding algebra is an example of the triple Lie algebras, which were introduced by Jacobson \cite{Jac51} in 1951. At the same time Green \cite{Green53} was introduced the
parafermionic algebra as an associative algebra, whose operators $f_i^\dagger, f_i  $  satisfy the ternary relations:
\[
\begin{array}{c}
\left[ \, f_k,  \,  \left[ f_\ell^\dagger  ,\, f_m \right] \right] =  2 \delta_{k \ell} f_m
\\
\left[ \,  f_k,  \,  \left[ f_\ell^\dagger  ,\, f_m^\dagger   \right] \right] =
 2 \delta_{k \ell} f_m^\dagger - 2 \delta_{k m} f_\ell ^\dagger
\\
\left[ \,  f_k,  \,  \left[ f_\ell  ,\, f_m \right] \right] = 0
\end{array}
\]

We call  parafermionic Poisson algebra the  Poisson algebra satisfying
the ternary relations:
\[
\left\{ x_i, \left\{ x_j, x_k \right\}_P \right\}_P
=\sum\limits_{m} c_{i;jk}^m x_m
\]
which is the classical Poisson analogue of the Lie triple algebra (\ref{eq:DefLieTriple}).

The quadratic parafermionic Poisson algebra is a Poisson algebra satisfying
the relations:
\[
\left\{ x_i, \left\{ x_j, x_k\right\}_P \right\}_P =
\sum\limits_{m,n}  d_{i;jk}^{mn} x_m x_n
 +\sum\limits_{m} c_{i;jk}^m x_m
\]
A classical  superintegrable system with quadratic integrals of motion on a two dimensional manifold possesses two functionally independent integrals of motion $A$ and $B$, which are in involution with the Hamiltonian $H$ of the system:
\[
\left\{H,\, A\right\}_P=0, \quad \left\{H,\, B\right\}_P=0
\]
the Poisson bracket $  \left\{A,\, B\right\}_P$ is different to zero and it is generally an integral of motion cubic in momenta, therefore it could not be in general  a linear combination of the integrals $H,\, A,\, B$.
Generally if we the Poisson brackets of the integrals of motion
 $\left\{ A,\left\{A,\,B\right\}_P\right\}_P$, $\left\{\left\{A,\,B\right\}_P ,B\right\}_P$ are not linear functions of the intergals of motion, therefore they don't close in a Lie Poisson algebra with three generators. If we consider all the nested Poisson brackets of the integrals of motion, generally they don't close in an Poisson Lie algebra structure.

All the known two dimensional superintegrable systems with quadratic integrals of motion the have a common structure \cite{BoDasKo94,Das00,Das01,DasYps06,DasTan07}:
\begin{equation}\label{eq:twodim}
\begin{array}{c}
\left\{H,\, A\right\}_P=0, \quad \left\{H,\, B\right\}_P=0,\quad
\left\{ A, B\right\}_{P}^2= 2 F(A,H,B)\\
\left\{ A , \left\{A,B\right\}_P\, \right\}_P = \displaystyle\frac{\partial F}{ \partial B} \;  , \;
\left\{ B ,  \left\{A,B\right\}_P\, \right\}_P =-\frac{\partial F}{ \partial A}
\end{array}
\end{equation}
where $F=F(A,B,H)$ is a cubic function of the integrals of motion
\begin{equation}\label{eq:structure_two_dim}
\begin{array}{rl}
F(A,B,H)=&\alpha A^3 + \beta B^3 +  \gamma A^2 B + \delta A B^2+
 \left(\epsilon_0 + \epsilon_1 H \right)  A^2  +
 \left(\zeta_0 + \zeta_1 H\right)  B^2  +\\
+&
\left(\eta_0+ \eta_1 H \right) A B  +
 \left(\theta_0 + \theta_1 H + \theta_2 H^2\right)  A+\\
+&\left( \kappa_0 + \kappa_1 H + \kappa_2 H^2\right) B + \left(\lambda_0+ \lambda_1 H + \lambda_2 H^2 + \lambda_3 H^3\right)
\end{array}
\end{equation}
where the greek letters are constants.

\section{Non degenerate three dimensional Kepler - Coulomb ternary parafermionic-like Poisson algebra}

The  non degenerate Kepler Coulomb Hamiltonian (\ref{eq:Non_degenerate}) has in three quadratic integrals which in agreement with \cite{Evans90} are:
\[
A_{1}=\frac{1}{2} J^2+\frac{k_{1}(x^2+y^2+z^2)}{x^2}+\frac{k_{2}(x^2+y^2+z^2)}{y^2}+\frac{k_3 (x^2+y^2+z^2)}{z^2}
\]
\[
A_{2}=\frac{1}{2} J_{3}^2+\frac{k_{1} (x^2+y^2)}{x^2}+\frac{k_{2} (x^2+y^2)}{y^2}
\]
\[
B_{2}=\frac{1}{2} J_{2}^2+\frac{k_{1} z^2}{x^2}+\frac{k_3 x^2}{z^2}
\]
where
\[
J_{1}=y p_{z}-z p_{y} ,\qquad J_{2}=z p_{x}-x p_{z} ,\qquad J_{3}= x p_{y}-y p_{x}, \qquad J^2=J_{1}^2+J_{2}^2+J_{3}^2
\]

The Coulomb potential differs from the other non degenerate potentials that described in \cite{KalKrMi07} since posses one integral of fourth order in addition to above, quadratic one which denoted by $ B_1$ and have the following form:
\[
B_{1}= \biggl (  J_1 p_{y}-J_2 p_{x}-2 z \Bigl ( \frac{-k}{2 \sqrt{x^2+y^2+z^2}}+\frac{k_1}{x^2}+\frac{k_2}{y^2}+\frac{k_3}{z^2} \Bigr )  \biggr )^2
+ \frac{2 k_3}{z^2} (x p_{x}+y p_{y}+z p_{z})^2
\]

One of the general results in \cite{KalKrMi07} is the so called "5 to 6" theorem:

\noindent\textbf{5$\to$6 Theorem}:
\textit{
Let $V$ be a nondegenerate potential (depending on 4 parameters) corresponding to
  a conformally flat space in 3 dimensions
  \[ ds^2= g(x,y,z) (dx^2+dy^2 +dz^2)\]
   that is superintegrable and there are 5 functionally
independent constants of the motion ${\mathcal L}=\{{ S}_\ell:\ell=1,\cdots 5\}$    There is always a $6$th quadratic integral ${S}_6$ that is
functionally dependent on $\mathcal L$, but linearly independent
}

\medskip

In case of Coulomb potential the sixth integral exist and is an integral fourth order in momenta as well the $B_1$ with general form given by the next expression:
\[
F= \biggl( -J_1 p_{z}+J_3 p_{x}-2 y \Bigl ( \frac{-k}{2 \sqrt{x^2+y^2+z^2}}+\frac{k_1}{x^2}+\frac{k_2}{y^2} +\frac{k_3}{z^2}\Bigr ) \biggr)^2+\frac{2
k_2}{y^2} (x p_{x}+y p_{y}+z p_{z})^2
\]
By studying all the known non degenerate potentials given by Kalnins, Kress, Miller \cite{KalKrMi07}, we can show that:
\\
\noindent{\textbf{Proposition:}}
\begin{textit}
{In the case of the non degenerate with quadratic integrals of motion, on a conformally flat
manifold, the integrals of motion satisfy a  parafermionic-like quadratic Poisson Algebra with 5  generators which described from the following:
\begin{equation}\label{eq:PoissonParaFermionicAlgebra}
\left\{ S_i,\, \left\{ S_j, \, S_k \right\}_{P} \right\}_{P}=
\sum\limits_{m n} d^{m,n}_{i;jk} S_m S_n + \sum\limits_{m} c^{m}_{i;jk} S_m
\end{equation}
}
\end{textit}

A detailed study of all the cases of non degenerate superintegrable systems on a flat space can be found in ref.\cite{TaDas09}. In all the cases (with one exception) the Poisson algebra of the integrals of motion is a ternary quadratic parafermionic-like Poisson algebra (\ref{eq:PoissonParaFermionicAlgebra}), which has a specific form.

In all the known cases (with only one exception) the non degenerate systems we can choose beyond the Hamiltonian $H$ four functionally independent integrals of motion $A_1,\, B_1, \, A_2, \, B_2$, and one additional quadratic integral of motion $F$, such that all the integrals of motion
are linearly independent. These integrals satisfy a Poisson parafermionic-like algebra (\ref{eq:PoissonParaFermionicAlgebra}).  The "special" form of the algebra defined by the integrals  $A_1,\, B_1, \, A_2, \, B_2$ is characterized by two cubic functions
\[
F_1=F_1\left(A_1,A_2,B_1,H\right) \quad
F_2=F_2\left(A_1,A_2,B_2,H\right)
\]
and satisfy the relations:
\begin{equation}\label{eq:SpecialStructure}
\begin{array}{c}
\left\{A_1,A_2\right\}_P=\left\{A_1,B_2\right\}_P= \left\{A_2,B_1\right\}_P=0,
\\
\begin{array}{l}
\left\{A_1,B_1\right\}_P^2= 2 F_1(A_1,A_2,H,B_1)=\mbox{cubic function}
\\
\left\{A_2,B_2\right\}_P^2= 2 F_2(A_1,A_2,H,B_2)=\mbox{cubic function}
\end{array}
\\
\left\{ A_i, \left\{A_i, B_i\right\}_P \right\}_P= \displaystyle \frac{\partial F_i}{\partial B_i} \; , \; \left\{ B_i, \left\{A_i, B_i\right\}_P \right\}_P=- \frac{\partial F_i}{\partial A_i}
\\
\left\{ \left\{A_1, B_1\right\}_P, B_2\right\}_P= \left\{ A_1,\left\{B_1, B_2\right\}_P\right\}_P, \quad \left\{\left\{A_2, B_2\right\}_P, B_1\right\}_P=- \left\{ A_2,\left\{B_1, B_2\right\}_P\right\}_P
\end{array}
\end{equation}
If we put
\[
C_1=\left\{ A_1,\, B_1\right\}_P,\quad C_2=\left\{ A_2,\, B_2\right\}_P,
\quad D=\left\{ B_1,\, B_2\right\}_P,
\]
the relations (\ref{eq:SpecialStructure}) imply the following ones:
\begin{equation}\label{eq:SpecialImplications}
\begin{array}{c}
\displaystyle\left\{C_1, B_2 \right\}_P C_1 - \frac{\partial F_1}{\partial A_2} C_2 - \frac{\partial F_1}{\partial B_1} D=\left\{C_2, B_1 \right\}_P C_2 - \displaystyle\frac{\partial F_2}{\partial A_1} C_1 + \displaystyle\frac{\partial F_2}{\partial B_2} D=0
\\
\left\{ C_1,C_2\right\}_P=
\displaystyle\frac{
\left|
\begin{array}{cc}
\left\{ A_1,D\right\}_P & - \displaystyle\frac{\partial F_1}{\partial A_2}\\
\displaystyle \frac{\partial F_2 }{\partial A_1}  & \left\{ A_2,D\right\}_P
\end{array}
 \right|
}{D} =
\displaystyle\frac{
\left|
\begin{array}{cc}
- \displaystyle\frac{\partial F_1}{\partial B_1}& \left\{ A_1,D\right\}_P \\
- \displaystyle\frac{\partial F_2}{\partial B_2}& \displaystyle \frac{\partial F_2 }{A_1}
\end{array}
 \right|
}{C_2}
=
\frac{
\left|
\begin{array}{cc}
- \displaystyle\frac{\partial F_1}{\partial A_2}&- \displaystyle \frac{\partial F_1}{\partial B_1}\\
\left\{ A_2,D\right\}_P&- \displaystyle \frac{\partial F_2}{\partial B_2}
\end{array}
 \right|
}{C_1}
\end{array}
\end{equation}
and
\[
\{C_1,C_2\}_P=\frac{\displaystyle \frac{\partial F_1}{\partial B_1} \frac{\partial F_2}{\partial A_1} C_1+\frac{\partial F_1}{\partial A_2}\frac{\partial F_2}{\partial B_2} C_2+\frac{\partial F_1}{\partial B_1}\frac{\partial F_2}{\partial B_2} D}{C_1 C_2}
\]

Schematically the structure of the above algebra is described by the following "$\Pi$" shape
\begin{equation}\label{eq:Pi_diagram}
  \xymatrix{
  A_1 \ar@{--}[d] \ar@{--}[r]
                & A_2 \ar@{--}[d]  \\
  B_2 \ar@{}[ur]
                & B_1 \ar@{}[ul]   }
\end{equation}
where with dashed line represented the vanishing of Poisson bracket whereas the other brackets between the integrals are non vanishing Poisson brackets.

It is  important to notice that the integrals $A_1,B_1$ satisfy a parafermionic-like quadratic Poisson algebra similar to the algebra as in two dimensional case (\ref{eq:twodim}). The corresponding structure function to the two dimensional one  (\ref{eq:structure_two_dim}) can be written as:
\begin{equation}\label{eq:structure_three_dim}
\begin{array}{rl}
F_1(A_1,B_1,H,A_2)=&\alpha_1 A_1^3 + \beta_1 B_1^3 +  \gamma_1 A_1^2 B_1 + \delta A_1 B_1^2+
 \left(\epsilon_{01} + \epsilon_{11} H  + \epsilon_{21}A_2\right)  A_1^2  +\\
 +&\left(\zeta_{01} + \zeta_{11} H + \zeta_{21} A_2\right)  B_1^2  +
\left(\eta_{01}+ \eta_{11} H+ \eta_{21} A_2 \right) A_1 B_1  +\\
 +&\left(\theta_{01} + \theta_{11} H + \theta_{21} H^2  + \theta_{31} A_2 + \theta_{41} A_2^2 + \theta_{51} A_2 H \right)  A_1+\\
+&\left( \kappa_{01} + \kappa_{11} H + \kappa_{21} H^2+ \kappa_{31} A_2 + \kappa_{41} A_2^2 + \kappa_{51} A_2 H\right) B_1 + \\
+&\lambda_{01}+ \lambda_{11} H + \lambda_{21} H^2 + \lambda_{31} H^3+
\lambda_{41} A_2 + \lambda_{51} A_2^2 + \lambda_{61} A_2^3+ \\
+& \lambda_{71} A_2 H + \lambda_{81} A_2^2 H +  \lambda_{91} A_2 H^2
\end{array}
\end{equation}

The pair $A_2,B_2$ forms also a parafermionic-like algebra with the corresponding structure function $F_2(A_2,B_2,H,A_1)$, which has a similar form as in (\ref{eq:structure_three_dim}).

The non degenerate Coulomb potential obey to the above parafermionic-like "$\Pi$" structure that characterize, almost all three dimensional superintegrable systems. Precisely the algebra needs some modifications due to the difference of the integrals order and to one extra symmetry that appears in system. In fact, the extra symmetries on a superintegrable system cause a number of changes to the default "$\Pi$" structure always respecting the "$\Pi$" shape. The extra symmetry that Coulomb potential have is $\{ B_1,F\}=0$ and the general structure can be schematically represented by the following figure:
 \begin{equation}\label{eq:Pi_diagram2}
  \xymatrix{
  A_2 \ar@{--}[d] \ar@{--}[r]
                & A_1 \ar@{--}[d]  \\
  B_1 \ar@{}[dr]
                & B_2 \ar@{--}[d] \\
                & F \ar@{}[u]  }
\end{equation}
The existence of one extra symmetry cause a basic difference to the general structure; one new subalgebra arising that expands in terms of $A_1,B_1,F$ integrals that also imply the presence of one extra function $F_3(A_1,B_2,F,H)$ which is a fourth order function. Particularly, the "special" form of the algebra defined by the integrals  $A_1,\, B_1, \, A_2, \, B_2, F$ is characterized by two fourth order functions and one cubic:
\[
F_1=F_1\left(A_1,A_2,B_1,H\right), \quad
F_2=F_2\left(A_1,A_2,B_2,H\right), \quad F_3=F_3\left(A_1,B_2,F,H\right)
\]
and satisfy the relations:
\begin{equation}\label{eq:SpecialStructure}
\begin{array}{c}
\left\{A_1,A_2\right\}_P=\left\{A_1,B_2\right\}_P= \left\{A_2,B_1\right\}_P=0,\left\{B2,F\right\}_P=0
\\
\begin{array}{l}
\left\{A_1,B_1\right\}_P^2= 2 F_1(A_1,A_2,H,B_1)=\mbox{fourth order function}
\\
\left\{A_2,B_2\right\}_P^2= 2 F_2(A_1,A_2,H,B_2)=\mbox{cubic function}
\\
\left\{A_1,F\right\}_P^2= 2 F_3(A_1,F,H,B_2)=\mbox{fourth order function}
\end{array}
\\
\left\{ A_i, \left\{A_i, B_i\right\}_P \right\}_P= \displaystyle \frac{\partial F_i}{\partial B_i} \; , \; \left\{ B_i, \left\{A_i, B_i\right\}_P \right\}_P=- \frac{\partial F_i}{\partial A_i}
\\
\left\{ A_1,\left\{A_1,F\right\}\right\}_P= \displaystyle \frac{\partial F_3}{\partial F} \; , \; \left\{ F,\left\{A_1,F\right\}\right\}_P= \displaystyle -\frac{\partial F_3}{\partial A_1}
\\
\left\{ \left\{A_1, B_1\right\}_P, B_2\right\}_P= \left\{ A_1,\left\{B_1, B_2\right\}_P\right\}_P, \quad \left\{\left\{A_2, B_2\right\}_P, B_1\right\}_P=- \left\{ A_2,\left\{B_1, B_2\right\}_P\right\}_P
\end{array}
\end{equation}

The structure functions of the above algebra are:
\[
\begin{array}{c}
F_{1}=4 k^2 A_{1} B_{1}-4 k^2 A_{2} B_{1}+4
    k_3 k^2 B_1-32  k_3 k^2 A_1 H-4 A_{1} B_{1}^2+16
   A_{1}^2 B_{1} H-\\-16 A_{1} A_{2} B_{1} H-64
    k_3 A_1^2 h^2+16  k_3 A_1 B_1 H-4 k_3 k^4
\end{array}
\]
\[
\begin{array}{c}
F_{2}=-4 k_1 A_{1}^2+4 A_{2} B_{2} A_{1}+8 k_1 A_{2}
   A_{1}+4  k_1 B_{2} A_{1}-4 k_2 B_{2}  A_{1}+8 k_1
   k_3 A_{1}-4 A_{2} B_{2}^2-\\-4  k_3 A_{2}
   B_{2}-4 A_{2}^2
   B_{2}-4 k_1 A_{2}^2 -4 k_1 A_{2} B_{2} + 4 k_2 A_{2}
   B_{2} -4 k_3 A_{2}^2 -4  k_1 k_3 B_{2}+8 k_2 k_3 A_{2} +\\+4  k_2 k_3 B_{2}
  +8 k_1 k_2 k_3-4 k_1 k_3^2-4 k_1^2 k_3-4 k_2^2 k_3
\end{array}
\]
\[
\begin{array}{c}
F_3=4 k^2 A_{1} F -4 k^2 B_{2} F -4 k_1 k^2 F +4 k_2
   k^2- 32 k_2 k^2 F A_{1} H -4 k_3 k^2 F -4 A_{1} F^2+\\+16
   A_{1}^2 F H-16 A_{1} B_{2} F H-16 k_1 A_{1} F H -64 k_2
   A_{1}^2 H^2 +16 k_2 A_{1} F H -16 k_3 A_{1} F H -4 k_2 k^4
\end{array}
\]
The full algebra is:
\[
\{\{A_1,B_1\},A_2\}=\{A_1,\{A_2,B_2\}\}=\{\{A_1,F\},B_2\}=0
\]
\[
\begin{split}
\{ \{A_1,B_1\},B_{1}\}= \frac{\partial F_{1}}{\partial A_{1}}  =-4 B_{1}^2+4 k^2 B_{1}+32 A_{1} H B_{1}-16 A_{2} H
   B_{1}+16 H k_3 B_{1}-\\-128 A_{1} H^2 k_3-32 H k^2 k_3
\end{split}
\]
\[
\{A_{1},\{A_1,B_1\}\}= \frac{\partial F_{1}}{\partial B_{1}} =16 H A_{1}^2+4 k^2 A_{1}-8 B_{1} A_{1}-16 A_{2} H
   A_{1}+16 H k_3 A_{1}-4 A_{2} k^2+4 k^2 k_3
\]
\[
\begin{split}
\{\{A_2,B_2\},B_{2}\}= \frac{\partial F_{2}}{\partial A_{2}} =-4 B_{2}^2+4 A_{1} B_{2}-8 A_{2} B_{2}-4 k_1
   B_{2}+4 k_2 B_{2}-4 k_3 B_{2}+8 A_{1} k_1-\\-8
   A_{2} k_1-8 A_{2} k_3+8 k_2 k_3
   \end{split}
\]
\[
\begin{split}
\{A_{2},\{A_2,B_2\}\}= \frac{\partial F_{2}}{\partial B_{2}}=-4 A_{2}^2+4 A_{1} A_{2}-8 B_{2} A_{2}-4 k_1
   A_{2}+4 k_2 A_{2}-4 k_3 A_{2}+4 A_{1} k_1-\\-4
   A_{1} k_2-4 k_1 k_3+4 k_2 k_3
\end{split}
\]
\[
\begin{array}{c}
\displaystyle \{\{F,A_1\},A_1\}=\frac{\partial F_3}{\partial F}=16 H A_{1}^2+4 k^2 A_{1}-8 F A_{1}-16 B_{2} H
   A_{1}-16 H k_1 A_{1}+16 H k_2 A_{1}-\\-16 H k_3 A_{1}-4
   B_{2} k^2-4 k^2 k_1+4 k^2 k_2-4 k^2 k_3
 \end{array}
   \]
\[
\begin{array}{c}
\displaystyle \{F,\{F,A_1\}\}=\frac{\partial F_3}{\partial A_1}=-4 F^2+4 k^2 F+32 A_{1} H F-16 B_{2} H F-16 H k_1 F+16 H k_2
   F-\\-16 H k_3 F-128 A_{1} H^2 k_2-32 H k^2 k_2
\end{array}
   \]
\[
\begin{array}{c}
\{A_1,\{B_1,B_2\}\}= \{\{A_1,B_1\},B_2\}= -16 H A_{1}^2-4 k^2 A_{1}+4 B_{1} A_{1}+4 F
   A_{1}+16 A_{2} H A_{1}-\\- 16 H k_3 A_{1}+4 A_{2}
   k^2+4 B_{1} B_{2}-4 A_{2} F+4 B_{1} k_1\\-4 B_{1}
   k_2-4 k^2 k_3+4 B_{1} k_3+4 F k_3
\end{array}
\]
\[
\begin{array}{c}
 \{\{A_2,B_2\},B_1\}= \{\{B_1,B_2\},A_2\}=16 H A_{1}^2+4 k^2 A_{1}-4 B_{1} A_{1}-4 F A_{1}-16
   A_{2} H A_{1}-\\-16 B_{2} H A_{1}-16 H k_3 A_{1}-4
   A_{2} k^2-4 B_{2} k^2+4 A_{2} B_{1}+4 B_{1}
   B_{2}+4 A_{2} F-\\-4 k^2 k_3+4 B_{1} k_3+4 F k_3
\end{array}
\]
\[
\begin{array}{c}
\{A_2,\{F,A_1\}\}=\{A_1,\{F,A_2\}\}=-16 H A_{1}^2-4 k^2 A_{1}+4 B_{1} A_{1}+4 F
   A_{1}+16 B_{2} H A_{1}+\\+16 H k_1 A_{1}-16 H k_2
   A_{1}+16 H k_3 A_{1}+4 B_{2} k^2-4 B_{1} B_{2}+4
   A_{2} F+4 k^2 k_1-4 B_{1} k_1-\\-4 k^2 k_2+4 B_{1} k_2+4
   k^2 k_3-4 B_{1} k_3-4 F k_3
\end{array}
\]
\[
\begin{array}{c}
\{\{ B_1, F\},A_1\}=-16 A_{1} B_{1} H+16 B_{1} B_{2} H+16 A_{1} F H-16
   A_{2} F H+\\+16 B_{1} k_1 H-16 B_{1} k_2 H+16 B_{1} k_3
   H+16 F k_3 H
\end{array}
\]
\[
\begin{array}{c}
\{\{A_1,B_1\},F\}=-64 A_{1}^2 H^2+64 A_{1} A_{2} H^2+64 A_{1} B_{2}
   H^2+64 A_{1} k_3 H^2-16 A_{1} k^2 H+\\+16 A_{2} k^2 H+16
   B_{2} k^2 H+16 A_{1} B_{1} H-16 B_{1} B_{2} H-16
   B_{1} k_1 H+\\+16 B_{1} k_2 H+16 k^2 k_3 H-16 B_{1} k_3 H-4
   B_{1} F
\end{array}
\]
\[
\begin{array}{c}
\{\{ A_1,F\},B_1\}=-4 (16 A_{1}^2 H^2-16 A_{1} A_{2} H^2-16 A_{1}
   B_{2} H^2-16 A_{1} k_3 H^2+\\+4 A_{1} k^2 H-4 A_{2} k^2
   H-4 B_{2} k^2 H-4 A_{1} F H+4 A_{2} F H-4 k^2 k_3 H-4 F
   k_3 H+B_{1} F)
\end{array}
\]
\[
\begin{array}{c}
\{ B_1,\{F,A_1\}\}=\{\{B_1,F\},A_2\}=-64 A_{1}^2 H^2+64 A_{1} A_{2} H^2+64 A_{1} B_{2}
   H^2+\\+64 A_{1} k_3 H^2-16 A_{1} k^2 H+16 A_{2} k^2 H+16
   B_{2} k^2 H+16 A_{1} B_{1} H-16 A_{2} B_{1} H-16
   B_{1} B_{2} H+\\+16 A_{1} F H-16 A_{2} F H+16 k^2 k_3
   H-16 B_{1} k_3 H-16 F k_3 H
\end{array}
\]
\[
\begin{array}{c}
\{ \{A_2,B_2\},F\}=\{B_2,M\}=-16 H A_{1}^2-4 k^2 A_{1}+4 B_{1} A_{1}+4 F
   A_{1}+16 A_{2} H A_{1}+\\+16 B_{2} H A_{1}+16 H k_3
   A_{1}+4 A_{2} k^2+4 B_{2} k^2-4 B_{1} B_{2}-4
   A_{2} F-4 B_{2} F-4 B_{1} k_1-4 B_{1} k_2+\\+4 k^2
   k_3-4 B_{1} k_3-4 F k_3
\end{array}
\]
\[
\begin{array}{c}
\{ \{B_1,B_2\},F\}=\{\{B_1 ,F\},B_2 \}=64 A_{1}^2 H^2-64 A_{1} A_{2} H^2-64 A_{1} B_{2}
   H^2-\\-64 A_{1} k_3 H^2+16 A_{1} k^2 H-16 A_{2} k^2 H-16
   B_{2} k^2 H-16 A_{1} B_{1} H+16 B_{1} B_{2} H-16
   A_{1} F H+\\+16 A_{2} F H+16 B_{2} F H+16 B_{1} k_1
   H+16 B_{1} k_2 H-16 k^2 k_3 H+16 B_{1} k_3 H+16 F k_3 H
\end{array}
\]
\[
\{\{ B_1,F\},F\}=64 A_{1} F H^2-64 A_{2} F H^2-64 B_{2} F H^2-128 B_{1}
   k_2 H^2-64 F k_3 H^2+16 B_{1} F H
\]
\[
\begin{array}{c}
\{\{ B_1,F\},B_1\}=-64 A_{1} B_{1} H^2+64 A_{2} B_{1} H^2+64 B_{1}
   B_{2} H^2+\\+64 B_{1} k_3 H^2+128 F k_3 H^2-16 B_{1} F H
\end{array}
\]
\[
\begin{array}{c}
\{ F,\{F,A_2\}\}=-4 F^2+4 k^2 F-4 B_{1} F+16 A_{1} H F+16 A_{2} H F-16 H k_1
   F+\\+16 H k_2 F-128 A_{1} H^2 k_2-32 H k^2 k_2+32 B_{1} H k_2
\end{array}
\]
\[
\begin{array}{c}
\{ A_2,\{F,A_2\}\}=4 A_{2} k^2-4 k_1 k^2+4 k_2 k^2-4 A_{2} B_{1}-8 A_{2}
   F+16 A_{1} A_{2} H+\\+4 B_{1} k_1-16 A_{1} H k_1-4
   B_{1} k_2+16 A_{1} H k_2
\end{array}
\]
\[
\begin{array}{c}
 \{ B_{1},\{B_1,B_2\}\}=-4 B_{1}^2+4 k^2 B_{1}-4 F B_{1}+16 A_{1} H
   B_{1}+16 B_{2} H B_{1}+\\+32 H k_3 B_{1}-128 A_{1}
   H^2 k_3-32 H k^2 k_3+32 F H k_3
\end{array}
\]
\[
\begin{array}{c}
\{ B_{2},\{B_1,B_2\}\}=-4 B_{2} k^2-8 k_3 k^2+8 B_{1} B_{2}+4 B_{2} F-16
   A_{1} B_{2} H+\\+8 B_{1} k_1+8 B_{1} k_3+8 F k_3-32
   A_{1} H k_3
\end{array}
\]

\section{Conclusions}\label{Conclusions}

The three dimensional non degenerate Kepler Coulomb potential \cite{{VerEvans08}} satisfy a ternary parafe\-rmionic-like fourth order Poisson algebra with quartic and quadratic integrals.
 The systems have three subalgebras forming a special "$\Pi"$ structure. Each subalgebra coreesponds to classical superintegrable system possessing two Hamiltonians.
The example of the non degenerate Kepler Coulomb system indicates that probably the known degenerate three dimensional potentials are related to non degerate systems with integrals of motion of order greater than two.

There is no results yet about the quantum  superintegrable systems as also there is not a compact general classification theory for three dimensional superintegrable potentials. The structure of the corresponding Poisson algebras for the degenarate systems is under investigation.

\vfill

\hrule

\bigskip

\textsf{MSC-class: 70h06, 17B63}

\end{document}